\documentclass{ws-p8-50x6-00}

\def\al{\alpha}  
\def\be{\beta}

\def\la{\lambda}

\def\Ga{\Gamma}
\def\Th{\Theta}

\newcommand{\ben}{\begin{equation}}
\newcommand{\een}{\end{equation}}
\renewcommand{\bea}{\begin{eqnarray}}
\renewcommand{\eea}{\end{eqnarray}}
\newcommand{\ba}{\begin{array}}
\newcommand{\ea}{\end{array}}
\newcommand{\bi}{\begin{itemize}}
\newcommand{\ei}{\end{itemize}}

\def\math{\mathsurround 0pt}
\def\oversim#1#2{\lower.5pt\vbox{\baselineskip0pt \lineskip-.5pt
        \ialign{$\math#1\hfil##\hfil$\crcr#2\crcr{\scriptstyle\sim}\crcr}}}

\def\gap{\mathrel{\mathpalette\oversim {\scriptstyle >}}}

\def\Re{\mathrm{Re}}
\def\Im{\mathrm{Im}}

\begin{document}

\title{Sphalerons with Two Higgs Doublets
\footnote{\lowercase{\uppercase{T}alk given at {\em \uppercase{S}trong and 
\uppercase{E}lectroweak 
\uppercase{M}atter}, \uppercase{M}arseille, 14-17 \uppercase{J}une 2000}}}

\author{Mark Hindmarsh and Jackie Grant}

\address{Centre for Theoretical Physics\\
University of Sussex\\
Falmer, Brighton BN1 9QJ\\
U.K.\\
{E-mail: m.b.hindmarsh@sussex.ac.uk, j.j.grant@sussex.ac.uk}}


\maketitle

\abstracts{
We report on work studying the properties of the sphaleron in 
models of the electroweak interactions with two Higgs doublets 
in as model-independent a way as possible: by
exploring the physical parameter space described by the masses and 
mixing angles of the Higgs particles.  If one of the Higgs particles 
is heavy, there can be several sphaleron solutions, distinguished by their
properties under parity and the behaviour of the Higgs field at the origin. 
In general, these solutions are not spherically symmetric, although the
departure from spherical symmetry is small.
SUSX-TH-01-001}

\section{Introduction}

One of the major unsolved problems in particle cosmology is accounting for the
baryon asymmetry of the Universe. This asymmetry is 
usually expressed in terms of the parameter
$\eta$, defined as the ratio between the baryon number density $n_B$ and the
entropy density $s$: $\eta = n_B/s \sim 10^{-10}$.  Sakharov~\cite{Sak67} laid
down the framework for any explanation: the theory of baryogenesis 
must contain B violation;
C and CP violation; and a departure from thermal equilibrium.  All these
conditions are met by the Standard Model~\cite{KuzRubSha85}
and its extensions, and so there is
considerable optimism that the origin of the baryon asymmetry can be found in
physics accessible at current and planned accelerators (see
\cite{Baryo} for reviews).

Current attention is focused on the Minimal Supersymmetric Standard Model, 
where there are many sources of CP violation over and above the CKM
matrix~\cite{CPMSSM},
and the phase transition can be first order for Higgs masses up to 120
GeV, providing the right-handed stop is very light and the
left-handed stop very massive~\cite{PhTranMSSM}.

B violation is provided by sphalerons~\cite{KliMan84}, at a rate 
$\Ga_s \simeq
\exp(-E_s(T)/T)$, where $E_s(T)$ is the energy of the
sphaleron at temperature $T$.  This rate must not be so large that the baryon
asymmetry is removed behind the bubble wall, and this condition can be
translated into a lower bound on the sphaleron mass
\( E_s(T_c)/T_c \gap 45.\)
Thus it is clear that successful baryogenesis requires a careful
calculation of the sphaleron properties.


Here we report on work on sphalerons in the two-doublet Higgs model (2DHM) in
which we study the properties of sphalerons in as general a set of
realistic models as possible.  In doing so we try to express parameter space in
terms of physical quantities: Higgs masses and mixing angles,
which helps us
avoid regions of parameter space which have already been ruled out by LEP.  

Previous work on sphalerons in 2DHMs
\cite{KasPecZha91,BacTinTom96,MorOakQui96,Kle98} has restricted either the
Higgs potential or the ansatz in some way. Our potential is restricted only by
a softly broken discrete symmetry imposed to minimize flavour-changing neutral
currents (FCNCs). Our ansatz is the most general spherically symmetric one, including
possible C and P violating field configurations~\cite{Yaf89}.

We firstly check our results against the existing literature, principally BTT
\cite{BacTinTom96}, who found a new P-violating ``relative winding'' (RW)
sphaleron, specific to multi-doublet models, albeit at $M_A=M_{H^\pm}=0$. 
This is distinguished from Yaffe's 
P-violating deformed sphaleron~\cite{Yaf89} or
``bisphaleron'' by a difference in the
behaviour of each of the two Higgs fields at the origin.
 We then reexamine the sphaleron in a more realistic part
of parameter space, where $M_A$ and $M_{H^\pm}$ are above their experimental
bounds.  
We reiterate the
point made in~\cite{GraHin98} that introducing Higgs sector CP violation makes a significant
difference to the sphaleron mass (between ten and fifteen percent),
and may significantly change  bounds on the
Higgs mass from electroweak baryogenesis.

\section{Two Higgs doublet electroweak theory}

The
most general quartic potential for 2DHMs has 14 parameters, only
one of which, the Higgs vacuum expectation value, $\upsilon$, is known.  However, we are
aided by the observation~\cite{HHG} that FCNCs can
be suppressed by imposing a softly-broken discrete symmetry 
$\phi_{1(2)} \rightarrow +(-) \phi_{1(2)}$, 
and results
in a potential with 10 real parameters.  One of these 
may be removed
by a phase redefinition of the fields; the vacuum configuration is then
entirely real, and CP violation is contained in one term 
\( 
2\chi_{2}\left(\Re( \phi_{1}^{\dagger}\phi_{2})-\frac{\upsilon_{1}\upsilon_{2}}{2}\right)
\Im(\phi_{1}^{\dagger}\phi_{2})\).
Ignoring couplings to other fields, when $\chi_2 = 0$ there is a discrete
symmetry
\(\phi_{\alpha}\rightarrow{-i}\sigma_{2}\phi_{\alpha}^{\ast}\),
which sends $\Im(\phi_{1}^{\dagger}\phi_{2})\to
-\Im(\phi_{1}^{\dagger}\phi_{2})$.  This can be identified as 
C invariance.  

Following~\cite{GraHin98} we
determine as many as possible of the nine parameters in the
potential from physical ones. The physical parameters at hand are the four
masses of the Higgs particles, the three mixing angles of the neutral Higgses,
one of which, $\theta_{CP}$, is the only CP violating physical parameter, ($\theta_{CP}$ 
mixes the CP even and CP odd neutral Higgs sector),
and $\upsilon$. 

We further check that the potential 
for these sets of physical parameters is always bounded from below.


\section{Sphaleron ansatz and numerical methods}

The most general static spherically symmetric ansatz is, in the radial gauge
\bea
\phi_{\alpha}=
(F_{\alpha}+{i}G_{\alpha}\hat{x}^a\sigma^{a})
\left(\begin{array}{c}0\\1\end{array}\right), ~~
W^a_i=
\left[\frac{(1+\beta)}{r}\varepsilon_{aim}\hat{x}_m
      +\frac{\alpha}{r}(\delta_{ai}-\hat{x}_a\hat{x}_i)\right]. \label{e:ansatz}
\eea
Here, the subscript $\alpha=1,2$, and $F_\al=a_\al+ib_\al$ and $G_\al=c_\al+id_\al$
are complex functions.
The boundary conditions can be most easily expressed in terms of the functions
$\chi$, $\Psi$, $H_\al$, $h_\al$, and $\Th_\al$, defined by
\bea
-\be + i \al  =  
\chi e^{ i\Psi}, ~~~
a_{\al} + i c_{\al} = 
H_{\al} e^{i \Th_{\al}}, ~~~
b_{\al} + i d_{\al} =  
h_{\al} e^ {i \Th_{\al}},
\eea
and one can show that as $r\to 0$, {either} $H_\al^2+h_\al^2  \to 0$ 
or $\Theta_1 \to \Psi/2 + n_1\pi$ and $\Theta_2  \to \Psi/2 + n_2 \pi$, ($n_1,n_2 \in {\mathbf Z}$).
These boundary conditions 
distinguish between the various types of
sphaleron solution:  the ordinary sphaleron has  $H_\al^2+h_\al^2  \to 0 $ as
$r\to 0$, the bisphaleron has non-vanishing Higgs fields with $n_1
= n_2$, and the RW sphaleron non-vanishing Higgs fields with
$n_1 \ne n_2$. These integers represent the winding of the Higgs field around
spheres of constant $|\phi_\al|$, although only their difference has any gauge
invariant meaning.

Note that the ansatz is potentially inconsistent, as $\Im(\phi_1^{\dagger}\phi_2) \propto \hat{x}_3$,
 a point which has not been noted before. In practice, the non-spherically
symmetric parts of the static energy functional, $E[f_A]$, contribute less than 1\% of the
total, and so we are justified in assuming the fields of the ansatz, $f_A$, are a function only of 
the radial co-ordinate $r$ and then integrating over $\hat{x}_3$.

We look for solutions to $E[f_A]$
using
a Newton method~\cite{Yaf89}
which is an efficient way of finding
extrema (and not just minima). The method can be briefly characterised as 
updating the fields $f_A$ by 
$\delta f_A$, given by the solution of 
${E}^{''} \delta f = - {E}^{'}$, where the primes denote functional differentialion 
with respect to $f_A$.
A particular advantage to using this method is that because we calculate ${E}^{''}$, it is straightforward to
get the curvature eigenvalues, and therefore to check that the solution really is the lowest energy
unstable solution.


\section{Results}

\begin{figure}[t]
\epsfxsize=6cm
\epsfbox{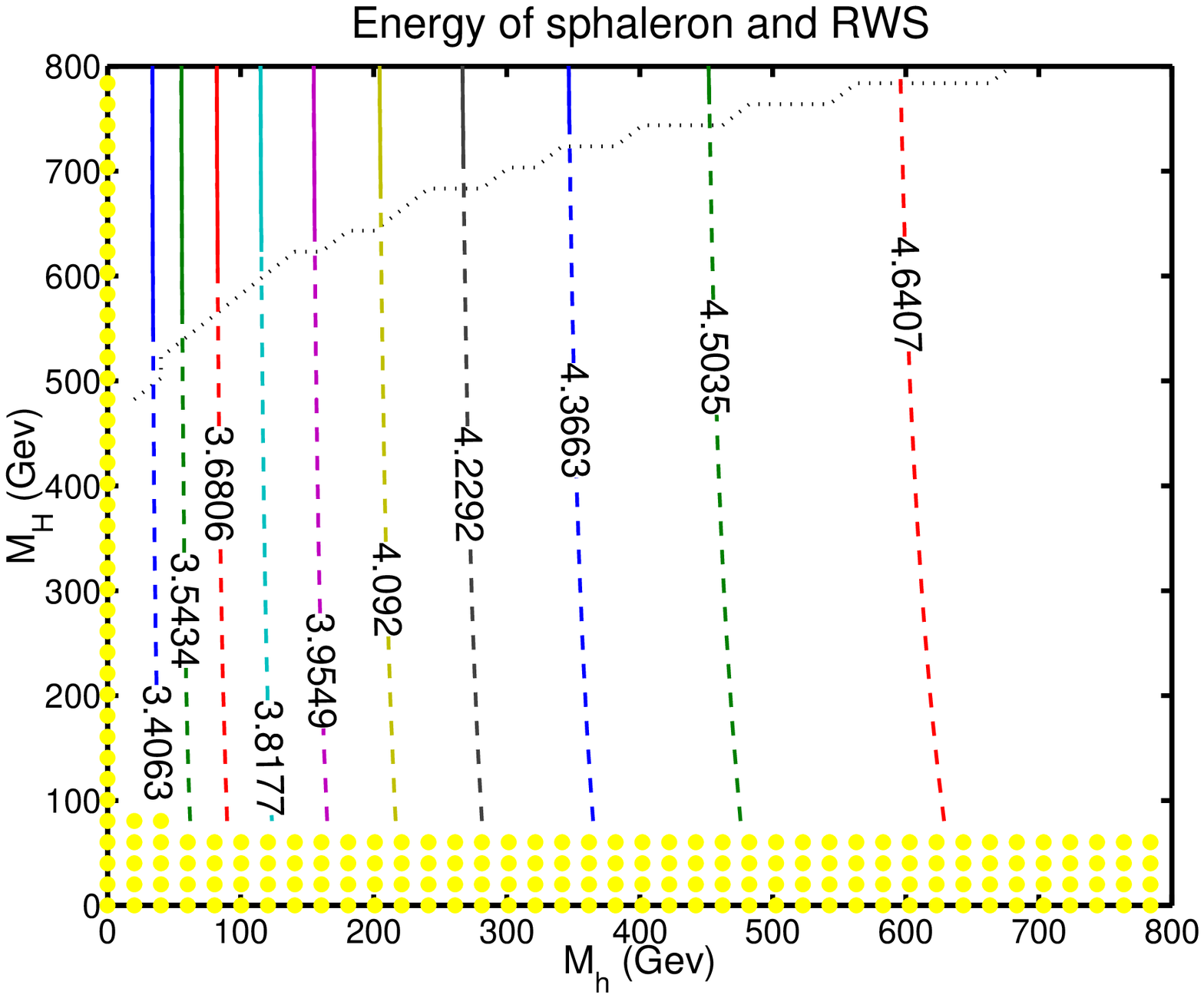}
\epsfxsize=6cm
\epsfbox{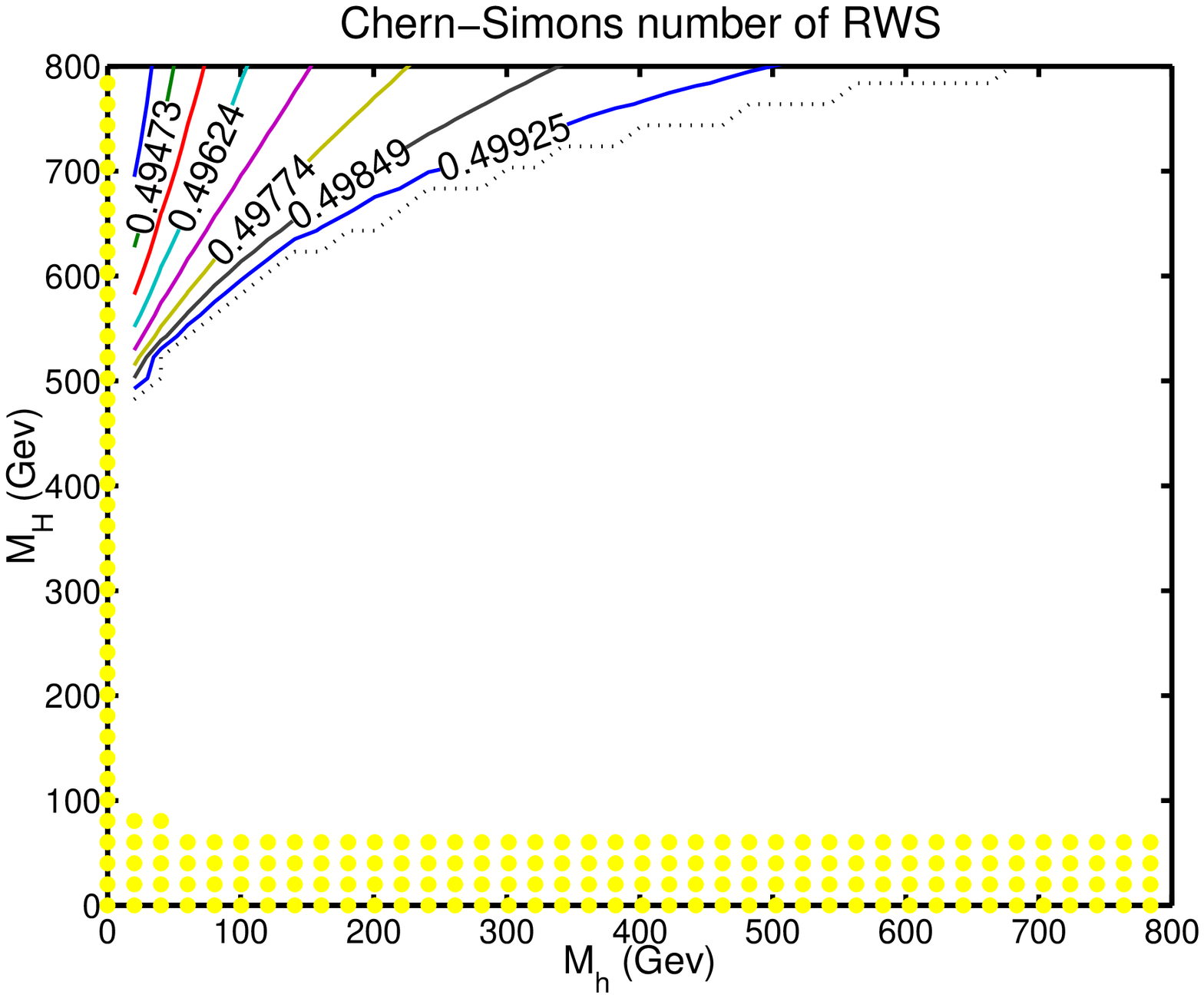}
\caption{\label{f:EC3} 
Contours of sphaleron energy ($M_W/\al_W$)
and Chern-Simons number. 
}
\end{figure}

\begin{figure}[t]
\epsfxsize=6cm
\epsfbox{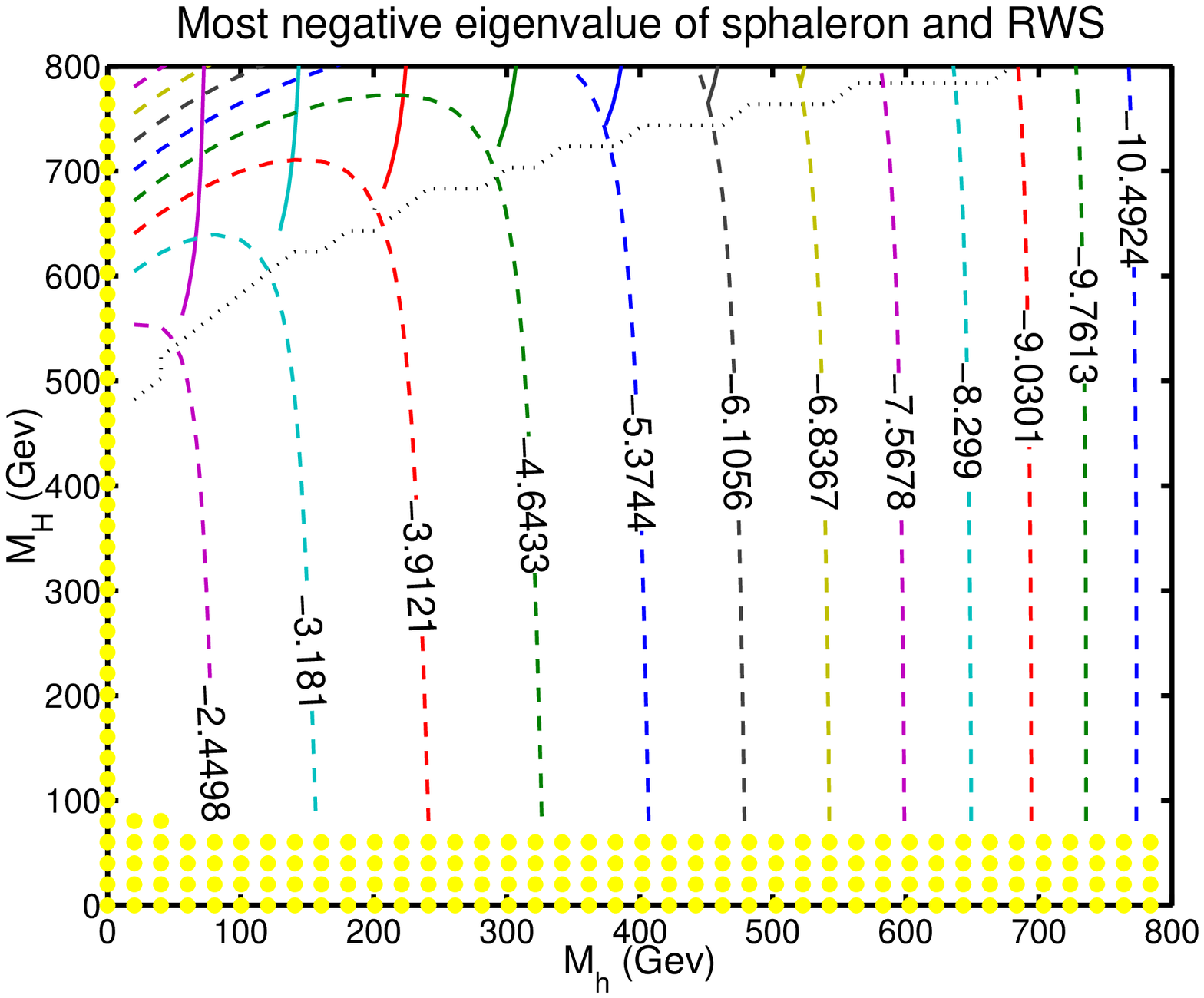}
\epsfxsize=6cm
\epsfbox{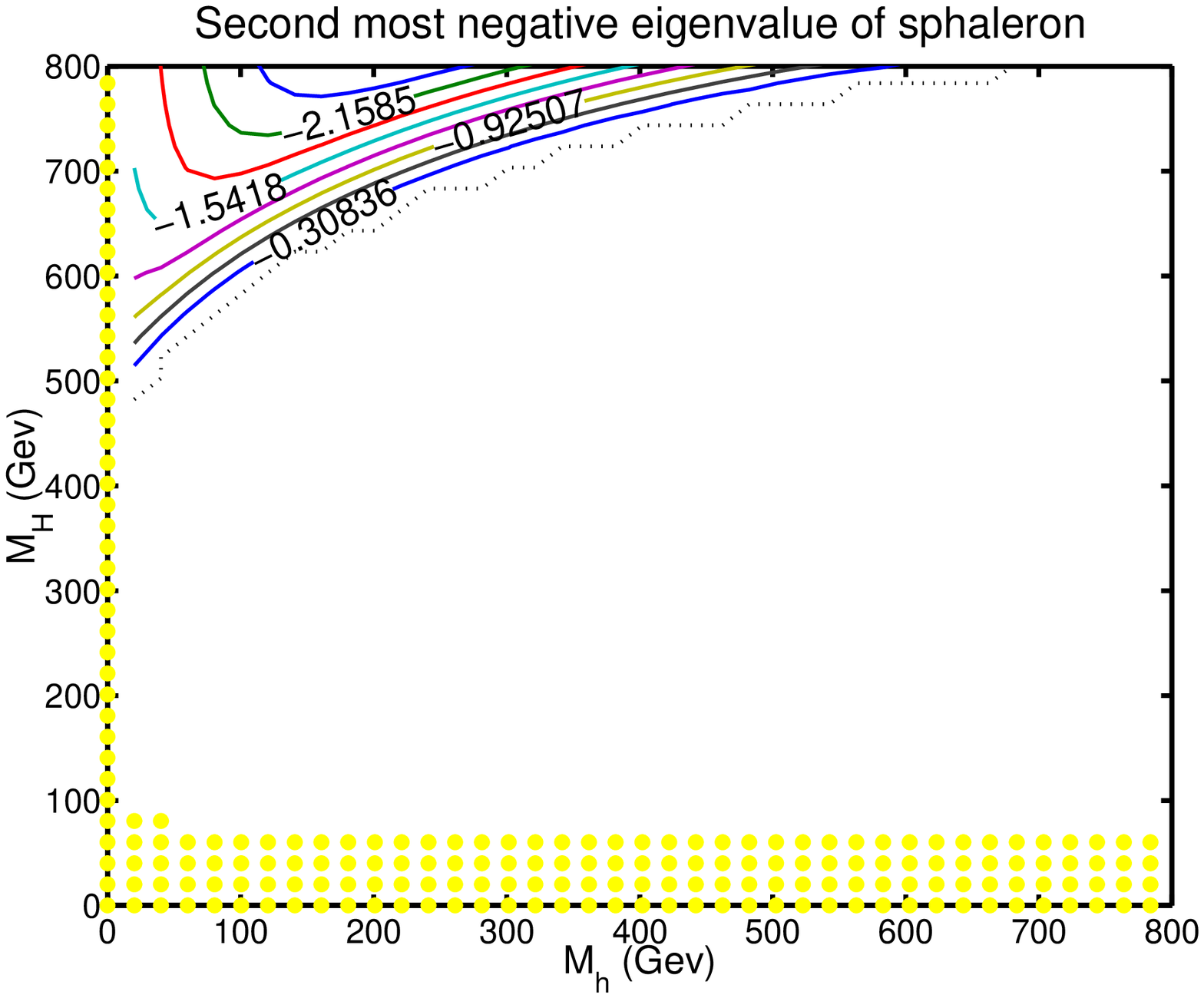}
\caption{\label{f:Eig3} 
Eigenvalues ($M_w$) of the sphaleron solutions as a function of 
the CP even Higgs masses $M_h$ and $M_H$.  There was no mixing, and 
$M_A = 241$ Gev, $M_{H^\pm}=161$ Gev, $\tan\be=6$, $\la_3=-0.05$. 
The solid lines in the 
most negative eigenvalue
plot represent the relative winding sphaleron, while the dashed lines
are for the ordinary sphaleron, for the dotted region the potential 
was unbounded from below.}
\end{figure}

We first checked our method and code against the results of BTT
\cite{BacTinTom96}, and Yaffe\cite{Yaf89}  finding agreement in the energy of better than 1 part in
$10^3$ for a wide range of parameters.  We also measured the Chern-Simons
numbers, $n_{CS}$, of the solutions that they discovered and determined that they were near
1/2, but not exactly 1/2 as with the sphaleron solution. Further they appeared 
in P conjugate pairs
with $n_{CS}$ of the pair adding to exactly one. 
We then looked at more realistic values of $M_A$ and $M_{H^\pm}$, with results
that are displayed in Figs.\ \ref{f:EC3} and \ref{f:Eig3}. 
Note first of all the well-known feature that the sphaleron mass
depends mainly on $M_h$.  Secondly, for increasing
$M_H$, the curvature matrix of the sphaleron
develops a second negative eigenvalue (Fig. \ref{f:Eig3}, right),
signalling the
appearance of a pair of RW sphalerons.  The lower of the two 
$n_{CS}$ is plotted on the right
Fig.\ \ref{f:EC3}.  
The departure from $n_{CS} = 1/2$ 
is 
small, as is the difference in energy between the RW and ordinary
sphalerons for the Higgs masses we examined, however the most negative curvature 
eigenvalue of the RW sphaleron can be double that of the sphaleron.
More detailed results and discussion of their significance 
are reserved for a future publication~\cite{GraHin00}.


\section*{Acknowledgments}
MH and JG are supported by PPARC.
This work was conducted on the SGI Origin platform using COSMOS
Consortium facilities, funded by HEFCE, PPARC and SGI.
We also acknowledge computing support from
the Sussex High Performance Computing Initiative.

\end{document}